\documentstyle[twocolumn,aps,prb]{revtex}
\begin{document}
\draft
\title{Ground-state properties of bosons in three- and 
 two-dimensional traps}
\author{Augusto Gonzalez$^{1,2}$\cite{adres} and Aurora 
  Perez$^2$\cite{aurora}}
\address{$^1$Departamento de Fisica, Universidad de Antioquia,
   AA 1226, Medellin, Colombia\\
   $^2$Instituto de Cibernetica, Matematica y Fisica Calle E 309,
   Vedado, Habana 4, Cuba}
\date{\today}

\maketitle

\begin{abstract}
We study trapped systems of bosons at zero temperature in three 
and two dimensions. Conditions are fulfilled for the application 
of Gross-Pitaevskii theory with a positive scattering length. 
Series expansions for ground-state properties are obtained in 
both the noninteracting and the strong-coupling (Thomas-Fermi) 
limits. From these expansions, analytic estimates are presented 
in the form of two-point Pad\'e approximants. We explicitly show
the approximants for the total energy per particle and the chemical 
potential.
\end{abstract}


Since the discovery of Bose-Einstein condensation in alkali-vapour
atoms \cite{BE}, trapped bosonic systems have attracted a lot of 
attention. In the experiments, the number of confined atoms, $N$, 
ranges between $10^4$ and $10^6$, whereas the ratio between the 
scattering length and the harmonic oscillator length is $a/a_{ho}
\sim 10^{-3}$, thus the conditions for the application of the 
Gross-Pitaevskii (GP) theory \cite{GP,GPS} are fulfilled. We assume
that the interaction between pairs of atoms is repulsive, i.e.
$a>0$, as for $^{87}$Rb. The temperature is taken to be zero, that 
is we are dealing with ground-state properties.

The GP equation for the condensate function, $\psi$, in three 
dimensions is written as

\begin{equation}
\left\{-\frac{\hbar^2}{2 m} \Delta+V_{ext}(\vec r)+
  g |\psi|^2(\vec r)-\mu\right\}\psi=0,
\end{equation}

\noindent
where $m$ is the mass of the atoms, $\mu$ -- the chemical potential, 
and $g=4\pi\hbar^2 a/m$. The condensate function satisfies the 
constraint

\begin{equation}
\int{\rm d}^3 r~|\psi|^2 =N.
\end{equation}

\noindent
In (1-2), we have neglected any effect coming from particles out 
of the condensate.
$V_{ext}(\vec r)$ is the external potential responsible for the 
confinement of the atoms. For simplicity, we will study an isotropic
trap, that is $V_{ext}=\frac{1}{2} m\omega^2 r^2$, and $\psi$ is 
a symmetric function $\psi(\vec r)=\psi(r)$.

A scaling of variables, $r\to a_{ho} r$, $\psi\to 
(N/a_{ho}^3)^{1/2} \psi$, $\mu\to \hbar\omega\mu$, reduces eqs. (1,2) 
to the dimensionless form

\begin{eqnarray}
\left\{ -\frac{1}{2} \Delta+\frac{1}{2} r^2+\tilde g |\psi|^2-
     \mu \right\} \psi=0,\\
\int{\rm d}^3 r~|\psi|^2 =1,
\end{eqnarray}

\noindent
and makes explicit that any scaled magnitude will depend only on the
variable $\tilde g=N g/(\hbar\omega a_{ho}^3)=4\pi N a/a_{ho}$, 
where $a_{ho}=\hbar^{1/2}/(m\omega)^{1/2}$. This scaling is 
preserved to the extent the GP equation remains valid. For example,
the $\omega$ can not be increased up to a value at which $a_{ho}$ 
becomes comparable to $a$.

In terms of the condensate function, the chemical potential is 
written as

\begin{equation}
\mu=\int {\rm d}^3 r\left\{ \psi^* (-\frac{1}{2} \Delta+
  \frac{1}{2} r^2)\psi+\tilde g |\psi|^4 \right\}.
\end{equation}

\noindent
It differs from the total energy per particle in half the Hartree 
energy

\begin{eqnarray}
E &=& \int {\rm d}^3 r\left\{ \psi^* (-\frac{1}{2} \Delta+
  \frac{1}{2} r^2)\psi+\frac{\tilde g}{2} |\psi|^4 \right\}
  \nonumber\\ 
  &=& \mu - \frac{\tilde g}{2} \int {\rm d}^3 r |\psi|^4.
\end{eqnarray}

We will consider formally that the variable $\tilde g$ ranges 
between zero (noninteracting bosons) and infinite (Thomas-Fermi
theory). In the $\tilde g\to 0$ limit, we may apply perturbation
theory, i.e. to look for $\psi$ and $\mu$ in the form

\begin{eqnarray}
\psi &=& \psi_0+\psi_1 \tilde g+\dots, \\
\mu &=& \mu_0+\mu_1 \tilde g+\dots.
\end{eqnarray}

The leading contributions are given by \cite{GPS,DS} $\psi_0=
\pi^{-3/4}e^{-r^2/2}$ , $\mu_0=3/2$. Next corrections are easily
obtained also, resulting in

\begin{eqnarray}
\psi_1 &=& -\sum_{n>0} \frac{\langle n|\psi_0^2|0\rangle }{2 n} 
 |n\rangle , \\
\mu_1 &=& \langle 0|\psi_0^2|0\rangle =\frac{1}{(2\pi)^{3/2}},
\end{eqnarray}

\noindent
where the $|n\rangle $ are three-dimensional harmonic oscillator 
states.

The series for $E$ is obtained from (6): $E_0=\mu_0$, whereas
$E_1=\mu_1/2$.

On the other hand, in the formal $\tilde g\to\infty$ limit, the 
kinetic energy may be neglected and the so called Thomas-Fermi (TF)
theory for bosons applies\cite{GPS,DS}. The condensate function 
is given by $\psi_{\infty}=\sqrt{(R^2-r^2)/(2\tilde g)}$, where
$R=(15\tilde g/(4\pi))^{1/5}$, and the chemical potential reads
$\mu_{\infty}=R^2/2$. The leading contribution to the energy is 
$E_{\infty}=5 R^2/14$.

The boundary layer near the condensate surface is responsible for 
the corrections to the TF theory\cite{FF}, leading to
contributions of the order of $1/R^2$,

\begin{eqnarray}
E &=& \frac{5}{14} R^2+\frac{5}{2 R^2} {\rm ln}(1.012~R)+\dots, \\
\mu &=& \frac{1}{2} R^2+\frac{3}{2 R^2} {\rm ln}(1.413~R)+\dots.
\end{eqnarray}

Taking together the weak-coupling and strong-coupling series, for 
anyone of the magnitudes $E$ and $\mu$ we can write

\begin{eqnarray}
\left. f(R)\right|_{R\to 0} &=& b_0+b_5 R^5+{\cal O}(R^{10}), \\
\left. \right|_{R\to\infty} &=& R^2 \left\{ a_0+\frac{a_4}{R^4} 
  {\rm ln}(A R)+{\cal O}(1/R^5)\right\}.
\end{eqnarray}

\noindent
The coefficients $b_0$, $b_5$, $a_0$ and $a_4$ are listed in Table 1.
Notice that the variable $R$ coincides with the condensate radius 
of the TF theory 
only in the $\tilde g\to\infty$ limit (differently from the notation 
used in Ref. [5]).

In most experimental conditions\cite{H97}, interactions are neither 
so weak to be considered in perturbation theory, nor so strong for the
TF theory to be valid. In the present paper, 
two-point Pad\'e approximants $\{P_{s,t}(R)\}$ are to be constructed
as analytic estimates to the magnitudes $f(R)$ over the entire range of 
variation of $R$. $s+1$ coefficients from the expansion (20) and $t+1$
from (21) are used to determine the coefficients in the approximant. 
Notice that in (13-14), we have at our disposal in total 15 coefficients
(many of which are equal to zero).

By construction, the approximants are asymptotically exact in both the 
$R \to 0$ and $R \to\infty$ limits. We will show that the error of the 
higher approximants is lower than a few percents at any $R$. 
A recent application
of the Pad\'e technique to obtain the ground-state energy of electrons 
in a parabolic quantum dot led to similar results\cite{2-5,2-210}.

As it is usual in Pad\'e-approximant techniques\cite{pade},  we shall 
show convergence of a sequence $\{P_{s,t}\}$, in which both $s$ and $t$ 
increase simultaneously. In the present problem, the sequence 
$\{P_{K+3,K}\}$ exhibits good convergence properties. The explicit 
expressions for the first nontrivial terms of this sequence are the 
following

\begin{eqnarray}
P_{5,2}(R) &=& b_0 + \frac{b_5 R^5} {1+q_1 R+q_3 R^3},\nonumber\\
           q_1 &=& b_0 b_5/a_0^2,~ q_3=b_5/a_0,\\ 
P_{6,3}(R) &=& b_0 + b_5 R^5 \frac{1+q_1 R}{1+q_1 R+\dots +q_4 R^4}
                   \nonumber\\
           q_3 &=& b_5/a_0,~ q_1=3 q_3/(2 a_0), \nonumber\\
           q_4 &=& b_5 q_1/a_0,~q_2 = 3 b_5 q_1/(2 a_0^2).
\end{eqnarray}

We show in Fig. 1 the relative difference between $P_{5,2}$ and
$P_{6,3}$ for the total energy per particle (the $K=2$ curve).
The maximum relative error of $P_{6,3}$ may be estimated from this
curve to be less than 7\% over the entire range of variation of $R$.
Analogous results are obtained for the chemical potential.

To build up the next approximant, we need the $\sim 1/R^2$ term 
in the high-$R$ series. However, this term contains a logarithmic
function of $R$. We may circumvent this problem\cite{IM} by 
constructing a (7,4) interpolant for the magnitude

\begin{eqnarray}
\left. \frac{1}{R} \frac{\rm d}{{\rm d}R} (R^2 f(R))
 \right|_{R\to 0} 
  &=& 2 b_0+ 7 b_5 R^5+ {\cal O}(R^{10}), \\
\left. \right|_{R\to\infty} &=& 4 a_0 R^2+\frac{a_4}{R^2} 
  +{\cal O}(1/R^3).
\end{eqnarray}

The series (17-18) are similar to (13-14) up to a redefinition of
coefficients: $\tilde b_0=2 b_0$, $\tilde b_5=7 b_5$, 
$\tilde a_0=4 a_0$, $\tilde a_4=a_4$. The approximant takes the form

\begin{eqnarray}
\tilde P_{7,4}(R) &=& \tilde b_0+\tilde b_5 R^5 
 \frac{1+Q_1 R+Q_2 R^2} {1+Q_1 R+\dots+Q_5 R^5}, \\
{\rm where}~Q_5 &=& \tilde b_5 Q_2/\tilde a_0, 
  Q_4=\tilde a_0 Q_2/\tilde b_0, \nonumber\\  
Q_3 &=& (\tilde a_0/Q_2+\tilde b_0) Q_5/\tilde a_0,\nonumber\\
Q_1 &=& Q_4 Q_2/Q_5, \nonumber
\end{eqnarray}

\noindent
and $Q_2$ is obtained from
$$\tilde a_0 Q_1-\tilde b_0 Q_3+\tilde a_4 Q_5=0.$$

The magnitude $f(R)$ is obtained by multiplying by $R$ and 
integrating the resulting expression . We will call it again the 
$P_{7,4}$ approximant,

\begin{equation}
P_{7,4}(R) = b_0+\frac{7 b_5}{R^2} \int_0^R {\rm d}x~x^6
  \frac{1+Q_1 x+Q_2 x^2}{1+Q_1 x+\dots+Q_5 x^5}.
\end{equation}

The integration over $x$ could be explicitly performed\cite{IM}, but 
a direct numerical integration of (20) is trivial. The coefficients
$Q_1,\dots,Q_5$ are listed in Table 1.

The relative difference between $P_{7,4}$ and $P_{6,3}$ is also 
shown in Fig. 1 (the $K=3$ curve). The maximum error of $P_{7,4}$ may
thus be estimated to be $\le 1.2 \%$. A similar result is obtained for 
$\mu$.

Next, we turn to the two-dimensional situation. We may think of 
an anisotropic three-dimensional trap in which $V_{ext}=
(\omega_{x,y}^2 r^2+\omega_z^2 z^2)/2$, where $r=\sqrt{x^2+y^2}$,
and $\omega_z/\omega_{x,y}>>1$, so that the motion of bosons in 
the $z$-direction is described by a gaussian of very small width.
Such highly anisotropic traps have been already constructed\cite
{2D}.

In eq. (1), we write $\Psi(r,z)=\chi(z)\psi(r)$, where $\chi(z)=
(m\omega_z/(\pi\hbar))^{1/4} e^{-m\omega_z z^2/(2\hbar)}$. 
Multiplying the 
equation by $\chi$ and integrating over $z$, the resulting equation
takes again the form (1), but the parameters entering it are 
$g=g_{3D} \int {\rm d}z~\chi^4=g_{3D}\sqrt{m\omega_z/(2\pi)}$, 
$\mu=\mu_{3D}-\hbar\omega_z/2$, and all the integrations will 
run over two-dimensional space. The reference level for any other 
magnitude with dimensions of energy will be $\hbar\omega_z/2$
also.

A scaling of variables, $r\to a_{ho} r$, $\psi\to 
(N/a_{ho}^2)^{1/2} \psi$, $\mu\to \hbar\omega\mu$, in which 
$\omega=\omega_{x,y}$, reduces the GP equation to the 
dimensionless form (3), where $\tilde g=
N g/(\hbar\omega a_{ho}^2)$. Notice that, differently from the 
3D case, $\tilde g$ does not depend 
on $\omega$. Thus, the dependence on $\omega$ of any magnitude 
can be obtained on purely dimensional grounds. For example,
any energy is exactly proportional to $\hbar\omega$. This is,
of course, valid to the extent the GP equation is valid.

In the $\tilde g\to 0$ limit, the solution is looked for as 
(7-8), leading to 

\begin{equation}
\psi_0=\pi^{-1/2}e^{-r^2/2},~~\mu_0=1,
\end{equation}

\begin{eqnarray}
\psi_1 &=& -\sum_{n>0} \frac{\langle n|\psi_0^2|0\rangle }{2 n} 
 |n\rangle,\\
\mu_1 &=& \langle 0|\psi_0^2|0\rangle =\frac{1}{2\pi}.
\end{eqnarray}

\noindent
Now, the $|n\rangle $ are two-dimensional harmonic oscillator 
states. The coefficients of the series for $E$ are, again, 
$E_0=\mu_0$, $E_1=\mu_1/2$.

On the other hand, in the $\tilde g\to\infty$ limit, TF 
theory leads to $\psi_{\infty}=\sqrt{(R^2-r^2)/(2\tilde g)}$, 
where $R=(4\tilde g/\pi)^{1/4}$, $\mu_{\infty}=R^2/2$, and
$E_{\infty}=R^2/3$. Corrections are again given by boundary
layer theory. We performed in 2D calculations similar to those
of Ref. [5]. The results are

\begin{eqnarray}
E &=& \frac{1}{3} R^2+\frac{4}{3 R^2} \ln (1.604~R)+\dots, \\
\mu &=& \frac{1}{2} R^2+\frac{2}{3 R^2} \ln (1.435~R)+\dots.
\end{eqnarray}

For anyone of the magnitudes $E$ and $\mu$, we have then

\begin{eqnarray}
\left. f(R)\right|_{R\to 0} &=& b_0+b_4 R^4+{\cal O}(R^8), \\
\left. \right|_{R\to\infty} &=& R^2 \left\{ a_0+\frac{a_4}{R^4} 
  {\rm ln}(A R)+{\cal O}(1/R^5)\right\}.
\end{eqnarray}

\noindent
The coefficients are listed in Table 2.

Pad\'e approximants are to be constructed from
(26-27). Notice that, in the approximation we are working, $f(R)$ 
contains only even powers of $R$ or $1/R$. Thus, the maximal power
in the Pad\'e should be even. In the sequence $\{P_{K+5,K}\}$, 
this leaves the approximants

\begin{eqnarray}
P_{5,0}(R) &=& b_0 + \frac{b_4 R^4} {1+(b_4/a_0) R^2},\\
P_{7,2}(R) &=& b_0 + b_4 R^4 \frac{1+q_2 R^2}{1+q_2 R^2+q_4 R^4}
                   \nonumber\\
           q_2 &=& b_4 a_0/(a_0^2-b_4),~ q_4=b_4 q_2/a_0. 
\end{eqnarray}

The relative difference between $P_{5,0}$ and
$P_{7,2}$ for the energy per particle in two dimensions 
is shown in Fig. 2 (the $K=0$ curve).
The maximum relative error of $P_{7,2}$ may be estimated to be 
$\le 9\%$ ($\le 10\%$ for $\mu$).

The next approximant in this sequence, the $P_{9,4}$, makes use not
only of the $a_4$ coefficient, but of the $b_8$ as well,

\begin{equation}
P_{9,4}(R) = b_0+\frac{6 b_4}{R^2} \int_0^R {\rm d}x~x^5
  \frac{1+Q_2 x^2+P_4 x^4}{1+Q_2 x^2+Q_4 x^4+Q_6 x^6},
\end{equation}

\noindent
where $P_4=(\tilde a_0/\tilde b_4) Q_6$, and $Q_2$, $Q_4$, $Q_6$ 
are obtained from the equations

\begin{eqnarray}
\tilde b_4 Q_4+\tilde b_8 &=& \tilde a_0 Q_6, \\
\tilde a_0 Q_4-\tilde b_0 Q_6 &=& \tilde b_4 Q_2, \\
\tilde a_0 Q_2-\tilde b_0 Q_4 &=& \tilde b_4 -\tilde a_4 Q_6, 
\end{eqnarray}

\noindent
The modified coefficients are $\tilde b_0=2 b_0$, $\tilde b_4=6 b_4$,
$\tilde b_8=10 b_8$, $\tilde a_0=4 a_0$, $\tilde a_4=a_4$. The numerical
values for $Q_2$, $Q_4$, $Q_6$, and $P_4$ are listed in Table 2.

The coefficient $b_8$ is obtained from second order perturbation 
theory. For the chemical potential the result is

\begin{equation}
\mu_2 =3 \langle \psi_0^3\psi_1\rangle = -\frac{3}{2} \sum_{n>0} 
   \frac{\langle n|\psi_0^2|0\rangle ^2}{ n}.
\end{equation}

The coefficient $E_2$ is obtained from (6), which may be 
rewritten in the $\tilde g\to 0$ limit as

\begin{equation}
E = \mu-\frac{\mu_1}{2} \tilde g-\frac{2\mu_2}{3} \tilde g^2+\dots.
\end{equation}

The relative difference between $P_{9,4}$ and $P_{7,2}$ is drawn
also in Fig. 2. It is labelled as the $K=2$ curve. From this curve, 
we may estimate the maximum relative error of the $P_{9,4}$
approximant to be less than $1.8 \%$ at any $R$.

A direct comparison with the numerical calculations of Ref. [12] 
for the energy per particle in two dimensions in the thermodynamic
($N\to\infty$) limit is presented in Fig. 3. The difference between 
both results is always below the predicted $1.8 \%$.

In conclusion, we obtained analytic Pad\'e estimations to 
ground-state properties of bosons in the Gross-Pitaevskii theory.
The approximants work with a small relative error ($< 2 \%$) at
any boson density. As examples, we explicitly found the 
approximants for the energy per particle and the chemical
potential in both three and two dimensions.

Pad\'e approximants for other magnitudes can be constructed in 
the same way. Different trap geometries could be considered also.
Finite systems, for which the GP formalism is no longer valid,
could be studied also because we can easily apply perturbation
theory near $g=0$, and use a Hartree approximation in the 
$g\to\infty$ limit. Finite temperatures could be equally well 
considered, in particular because of the scaling of thermodynamic
magnitudes\cite{GPS2}. All these analytic estimations could be 
of great value for the experimental groups.

\acknowledgements 

The authors acknowledge support from the Colombian
Institute for Science and Technology (COLCIENCIAS) and from the 
Research Grant Programme of the Third World Academy of Sciences
(TWAS). Both authors are associate members of the International 
Center for Theoretical Physics (ICTP, Trieste).

\begin{figure}
\caption{Relative differences between consecutive
  approximants for the energy in three dimensions.}
\label{fig1}
\end{figure}

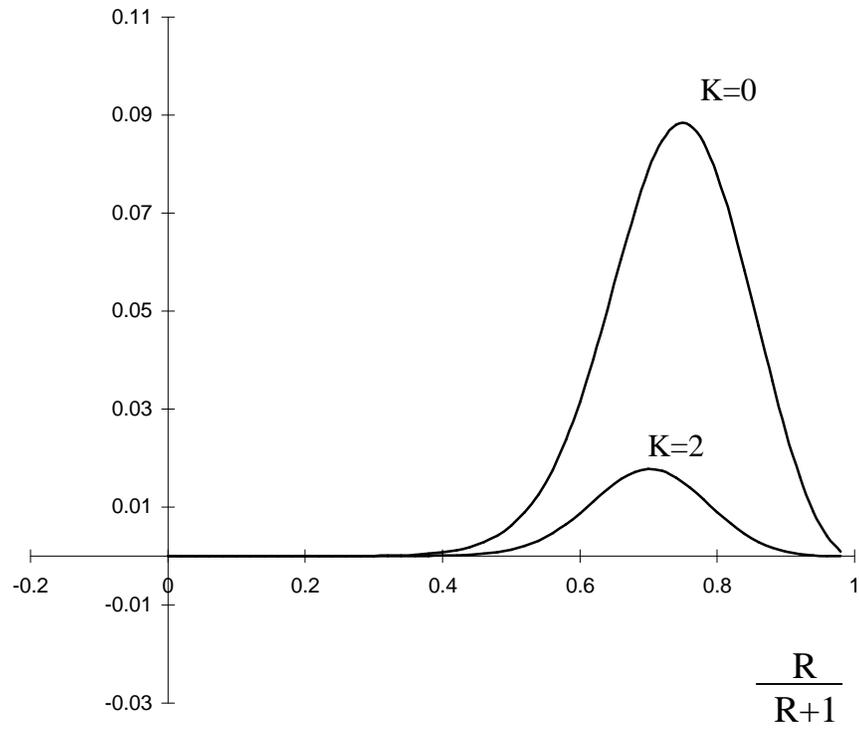
\begin{figure}
\caption{The same as in Fig. 1, but in two dimensions.}
\label{fig2}
\end{figure}

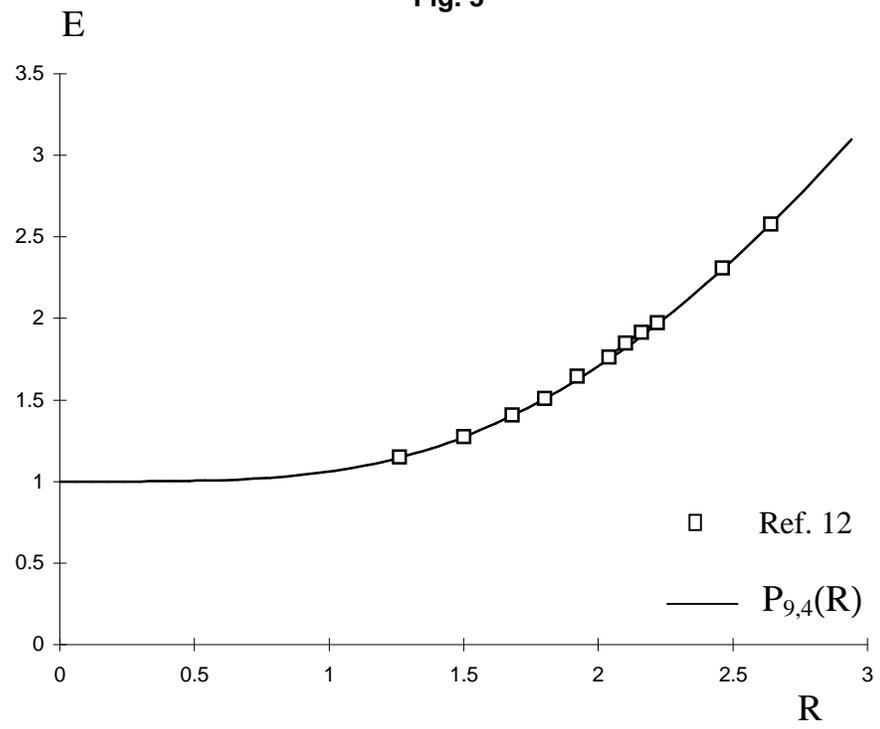
\begin{figure}
\caption{Comparison between the $P_{9,4}$ approximant and 
    the numerical calculations of Ref. 12 for the energy 
    per particle in two dimensions.}
\label{fig4}
\end{figure}

\begin{table}
\caption{Coefficients in 3D.}
\label{tab1}
\begin{tabular}{|l|l|l|}
  & $E$ & $\mu$ \\
\hline\hline
$b_0$ & 3/2                  & 3/2\\
$b_5$ & $1/(15 \sqrt{2\pi})$ & $2/(15 \sqrt{2\pi})$\\
$a_0$ & 5/14                 & 1/2\\
$a_4$ & 5/2                  & 3/2\\
$Q_1$ & 0.302359             & 0.30288\\
$Q_2$ & 0.082748             & 0.084582\\
$Q_3$ & 0.152967             & 0.209793\\
$Q_4$ & 0.039404             & 0.056388\\
$Q_5$ & 0.010784             & 0.015747\\
\end{tabular}
\end{table}

\begin{table}
\caption{Coefficients in 2D.}
\label{tab2}
\begin{tabular}{|l|l|l|}
  & $E$ & $\mu$ \\
\hline\hline
$b_0$ & 1         & 1\\
$b_4$ & 1/16      & $1/8$\\
$b_8$ & -0.001124 & -0.003371\\
$a_0$ & 1/3       & 1/2\\
$a_4$ & 4/3       & 2/3\\
$Q_2$ & 0.646107  & 0.751621\\
$Q_4$ & 0.292455  & 0.424002\\
$Q_6$ & 0.073825  & 0.142145\\
$P_4$ & 0.262488  & 0.379052\\
\end{tabular}
\end{table}

\end{document}